\documentclass[12pt,preprint]{aastex}
\usepackage{lscape,afterpage}
\usepackage{graphicx}

\begin{document}


\shorttitle{Second GALEX Ultraviolet Variability Catalog}
\shortauthors{Wheatley et al.}



\title{THE  SECOND {\it GALEX}  ULTRAVIOLET VARIABILITY (GUVV-2) CATALOG}


\author{
Jonathan M. Wheatley,\altaffilmark{1}
Barry Y. Welsh,\altaffilmark{1}
and Stanley E. Browne\altaffilmark{1}}

\altaffiltext{1}{Experimental Astrophysics Group, Space Sciences Laboratory, University of California, 7 Gauss Way, Berkeley, CA 94720;  wheat@ssl.berkeley.edu, bwelsh@ssl.berkeley.edu, seb@ssl.berkeley.edu}



\begin{abstract}
 We present the second
 Galaxy Evolution Explorer ($\it GALEX $) Ultraviolet Variability (GUVV-2) Catalog that
 contains information on
410 newly discovered time-variable sources gained through simultaneous near
(NUV 1750~-~2750~\AA) and far (FUV 1350~-~1750~\AA) ultraviolet
photometric observations. Source variability was determined by comparing the 
NUV and/or FUV fluxes derived from orbital exposures recorded during a series of multiple
observational visits to 169 GALEX fields on the sky.
These sources, which were contained within a sky-area of 161 deg$^{2}$, varied
on average by $\Delta$NUV = 0.6 mag and $\Delta$FUV = 0.9 mag
during these observations. Of the 114 variable sources in the catalog with previously known identifications, $\sim$ 67$\%$ can be categorized
as being active galaxies (QSO's, Seyfert 1 or BL Lac objects). The next largest groups of UV variables
are RR Lyrae stars, X-ray sources and novae. 

By using a combination of UV and visible color-color plots we have been able to
tentatively identify 36 possible RR Lyrae and/or $\delta$ Scuti
type stars, as well as 35 probable AGN's, many of which may be previously
unidentified QSO's or  blazars. Finally, we show data for 3 particular
variable objects:  the contact binary system of SDSS J141818.97+525006.7,
 the eclipsing dwarf nova system of  IY UMa
and the highly variable unidentified source SDSS J104325.06+563258.1.

\end{abstract}




\section{Introduction}
The NASA Galaxy Evolution Explorer
($\it GALEX$) satellite, launched on 2003 April 28th,
is currently making  imaging photometric observations of the sky
in two ultraviolet bands (NUV 1750 - 2750\AA\ , FUV 1350 - 1750\AA).
A summary of the main scientific findings obtained during the first year
of its on-orbit operation can be found in \citep{mar05} and references
therein.
Repeated (i.e. multi-orbit) observations of selected areas of the sky
are made in the Deep Imaging Survey (DIS) \citep{mor05, mor07} and in certain Guest 
Investigator (GI) observations.
This observing strategy requires
making repeated visits to the same position
on the sky with the 1.2$^{\circ}$ instrument field of view, thus enabling numerous astronomical sources to have their FUV and NUV
fluxes to be determined at many different epochs. From these observations
it thus possible to detect variable ultraviolet sources, 
many of which exhibit much larger amplitudes of variation in the
ultraviolet region than that typically found at visible wavelengths. A list of
84 time-variable UV sources discovered during the first 14 months of the
$\it GALEX$ mission (i.e. 2003 June to 2004 August) has been presented by
\cite{welsh05} in version 1.0 of the $\it GALEX$ Ultraviolet Variability (GUVV)
Catalog. The great majority of these 84 UV variable objects were found to be either RR Lyrae
or dMe flare stars, together with a few Delta Scuti and X-ray variable objects.
Since the publication of the first GUVV catalog several papers have been published on the
more extreme cases of UV source variability, such as the $\it GALEX$ high time-resolution
($<$ 1 sec) observations of flares
on 4 nearby dMe stars \citep{welsh06} and the large amplitude UV variations found for
the RR Lyrae star ROTSE-I J143753.84+345924.8 \citep{wheat05}. We note that the observation
of flux variability in astronomical objects can provide important constraints on the physical
processes responsible for the observed emission, especially in the cases of X-ray/UV emission
from highly energetic sources such as  AGN and black hole candidates.

In this Paper we report on the analysis of  new $\it GALEX$ observations
of 169 sky-fields performed during the 2003 June  to 2006 June timeframe.
Each sky-field was observed at least 10 times, each for a period
of  1200 sec to 1700 sec.  We report on 410 newly discovered UV variable sources
which will subsequently require  follow-up observations
in other wavelength bands in order to fully describe the true physical nature associated with
the variability of each of the listed sources.

\section{Observations and Data Analysis}
We have used the $\it GALEX$ FUV and NUV-band photometric data catalogs
compiled during the period
2003 June to 2006 June, which reside in the Multi-Misison
Archive at the Space Telescope Science Institute (MAST).
Only the $\it GALEX$ fields observed on 10 or more occasions (and
with exposures of more than 200 sec) were
included in our present analysis. A list of each sky-field, together with the
number of separate observations (or `exposures'), is given in Table 1 and the distribution
of these fields on the sky is shown in Figure 1.  Since the prime
scientific reason for selecting these fields was to carry out $\it GALEX$ UV observations
of galaxies and/or galaxy clusters, it is not surprising that
the vast majority of
these sky-fields are located away from the galactic plane such that saturation
of the detectors due to overly-bright stellar sources and the effects of
interstellar absorption in the galactic plane are both minimized. 

The  data files for each exposure of the 169 1.2$^{\circ}$ diameter 
sky-fields contain photon events that
have been processed using the standard $\it GALEX$ Data Analysis Pipeline operated
at the Caltech Science Operations Center (Pasadena, CA). This
pipeline ingests time-tagged
photon lists, instrument and spacecraft housekeeping data and
satellite pointing aspect information \citep{mor07}. The
data pipeline then uses a source detection
algorithm called SExtractor \citep{bertin96} to produce a catalog of 
source positions on the sky with corresponding FUV and NUV source magnitudes
calculated for
each observational visit. Source magnitudes derived from use of
a fixed 12 arcsec diameter aperture with SExtractor were utilized in this study, since these are
more appropriate for isolating flux from stellar sources rather than galaxies. The source detections
within each exposure at each sky-field
position on the sky appear as source lists in the $\it GALEX$ archival database
of the MAST.

Comparison software was then run on the MAST source list
catalog to reveal objects that we deemed as being
time-variable. This procedure
involved the following 4 steps that were applied
to each of the individual exposures of every source detected within each of the 169 
sky-fields: (1) Firstly, only objects brighter
than NUV magnitude m$_{NUV} = 21.0$, that were also located
within 0.55 degrees of the center of each $\it GALEX$ field were selected. 
Such a choice lessens the influence of both degraded
image resolution and `detector edge effects' that can potentially cause
artifacts in the data. The choice of magnitude limit was based on
ultimately being able
to maximize the number of truly variable sources while minimizing a far larger number
of spuriously variable fainter sources.
Targets within this selected area that 
were flagged by the $\it GALEX$ pipeline as being potential
 image artifacts in the MAST data-set were also
rejected. The result from this search was a list of sources,
sky positions and values of NUV (and FUV) magnitude that were found
within each exposure associated with each of the 169 sky-fields.
(2) A comparison was then
performed on all values of m$_{NUV}$ that were
obtained for each source from the various exposures of each of
the sky-fields. Since
the listed source positions each have an inherent measurement error of
$\sim$ $\pm$ 3 arc sec \citep{mor05}, we restricted the comparison to sources
listed within a 10 arc sec diameter error circle of the nominal source position. Only sources
that varied by $>$ 0.6 NUV magnitudes within this
error circle were selected as being potentially
time-variable. This variability criterion corresponds to a  $>$ 3$\sigma$ change
in the photon statistical error assigned to each source by the $\it GALEX$ data pipeline.
However, we note that for faint sources the GALEX systematic error (which is due to
the combined effects of sky-subtraction, flat fielding and
scattered light subtraction) is far larger
than the photon statistical error \citep{mor05}. Thus, the choice of $>$ 0.6 NUV mag. significantly minimizes
the number of false variability detections.
We note that the photometric stability of $\it GALEX$ assessed during
the 2003 - 2006 period is $<$ 0.1 magnitudes \citep{mor07}.
(3) When available, a comparison was also made with the
corresponding FUV magnitudes, m$_{FUV}$, for each exposure of the
sources previously identified in Step (2). If the selected source $\it also$ varied
by $>$ 0.4 FUV magnitudes, then this was deemed a confirmation of true
source variability. In cases where no FUV data were available, Step (3)
was omitted. (4) The previous two steps resulted in a list of
NUV (and/or FUV) source magnitudes with their associated sky positions recorded
at a given epoch for each exposure of the various sky-fields. 
Each of these source lists
was then plotted (as epoch versus magnitude) and the individual light-curves
visually checked
for the plausibility of astronomical source variability. The images of individual GALEX visits were
also visually checked as a final verification that the source flux variation
was not caused by some image anomaly. A typical anomaly in GALEX
images is scattered light leakage caused by bright objects positioned just outside of
the instrument field-of-view. In Figure 2 we show two GALEX images recorded
for the ELAIS-N1 field, taken 2.5 months apart. The fuzzy horseshoe-like feature to the left
of center of the 2005 May 3rd image is due to scattered light and is seen to re-appear
in the 2005 July 24th image with a greatly reduced intensity at a 
different angle to the two central bright objects, due to the different roll angle used in this
pointing of the GALEX instrument. 
Finally, we also ran a search for FUV source variability that was not accompanied by any
appreciable NUV
variability. This entailed running Steps (1) and (2) within the constraints of
m$_{FUV}$ $<$ 22.0 and an FUV magnitude variation of $>$ 1.0 magnitude.

The list of 410 resultant time-variable UV sources is given in Table 2. In column (1),
following the format of the GUVV-1 catalog \citep{welsh05},
we list a unique identifier for each
source that contains its $\it GALEX$ mean right ascension (J2000.0) in hours, minutes
and decimal seconds and its corresponding declination (J2000.0) in degrees, arcminutes
and decimal seconds. In column (2) we list, when available, the USNO-B1.0
all-sky catalog designation in $\it italics$ \citep{mon03} or the Sloan Digital Sky Survey
DR6 designation in Roman typeface \citep{adel07}.
For objects that appear in the
SIMBAD on-line astronomical catalog, in column (3) we list their source identification.
In column (4) we list, when known, the most likely 
astronomical source-type for the source. Criteria used to
make this latter determination were generally varied, but (for the
brighter sources) are mainly based
on either their Simbad catalog identifications or on
inspection of their $\it GALEX$ UV light-curve data (gained from Step 4 of the data search).
The majority of these variable sources have no known associated source
type, but when known we list source types in  the following
11 categories: QSO's (quasar stellar objects), Sy1 (Seyfert 1 galaxies),
AGN (active galactic nucleii), BLL (BL Lacs), Gal (ordinary galaxies), RR* (RR Lyrae stars), X (X-ray sources), V* (variable star),  ** (double stars), Nov (novae and dwarf novae) and Fl* (dMe flare stars).

In column (5) we list the name of the $\it GALEX$ sky-field in which
the variable source was found.
In column (6) we list the total number of
detections (NUV$_{det}$) of the variable source within the associated set of NUV observations
of the $\it GALEX$ sky-field.
Column (8) lists the
maximum observed NUV magnitude (NUV$_{max}$) for the source (measured
in a single exposure) and column (9)
lists the variation between the corresponding maximum
and minimum NUV magnitudes (i.e. $\Delta$NUV).
Similarly, columns (10) - (13) list the equivalent number of detections (FUV$_{det}$),  maximum magnitude (FUV$_{max}$) and variation in magnitude ($\Delta$FUV) for the FUV channel.
We emphasize that the non-detection of
a source previously observed in both (or one) of the two UV-bands
can be attributed to either intrinsic variability (i.e. an astrophysical effect) or being due to
 one of the detectors having been turned off
during a particular observation for instrument safety reasons.
Finally, in columns (14) - (16) we list (in Roman typeface)
the respective $\it g$, $\it r$ and $\it i$ photometric PSF magnitudes as recorded by the
Sloan Digital Sky Survey (SDSS) catalog \citep{adel07} for the source designation
listed in column (2). Finally, we note that this new GUVV-2 catalog also includes some of the variable
sources listed in the GUVV-1 catalog, often with different magnitude changes.
This is due to either a greater number of observations per source being presently
available for analysis, or the improvement in the source detection algorithms of
Version 5.1 of the $\it GALEX$ data pipeline software. For sources that have
no or uncertain SDSS photometric magnitudes, we list the
available USNO-B B, R and I magnitudes in $\it italics$.

\section{Discussion}
In this section we analyze some of the statistical properties of the 410 UV variable sources
listed in Table 2. The present sample of variable sources has been discovered
through observations covering a total area of $\sim$ 161 deg$^{2}$ on the sky, with the vast majority
of the fields
being located at galactic latitudes well away from the galactic plane. The previous GUVV-1 study of
\cite{welsh05} detected 84 variable sources contained within a far
larger area on the sky of $\sim$ 3000 deg$^{2}$. The larger area covered was due to
the use of many All-Sky Imaging Survey (AIS) fields recorded during the initial survey
phase of the $\it GALEX$ mission. The current  GUVV-2 detection rate is $\sim$ 2.6 variable UV sources
per deg$^{2}$. This can be directly compared with a GUVV-1 detection rate from DIS
pointings of $\sim$ 1.2
variable sources per deg$^{2}$ gained from observations of 15 deg$^{2}$.
The increase in the present detection rate is due to improved variability search software
and the generally longer time series of exposures over which variability
could be assessed within the presently observed fields.

\subsection{Variability Statistics}
In Figure 3 we plot the respective maximum changes in
UV magnitude, NUV$_{max}$ and FUV$_{max}$, as a function of
the number of sources exhibiting such variability. Although
we see a maximum number of 
NUV variable sources with $\Delta$NUV = 0.6 mag, the sharp cut-on in this source distribution
is due to our definition of NUV variability as being greater than 0.6 mag. Unfortunately selecting
an NUV magnitude variability less than this value results in a very large number of false
detections, particularly for sources with NUV magnitudes fainter than NUV = 19.0.
In the case of (associated) FUV source variability we see a maximum number of sources
at $\Delta$FUV = 0.9 mag. In Figure 4 we plot values of NUV$_{max}$ versus
FUV$_{max}$ for these sources and find that the great majority of targets
lie within $\pm$ 1 mag of a straight line of slope +0.91. This is very similar
to the results found for the far smaller sample of UV variable sources
in the  GUVV-1 catalog \citep{welsh05}.  We note the large number of sources
with peak FUV and NUV magnitudes $>$ 20.0 that deviate more than this value
from the best-fit line, and this is due to the larger measurement errors associated
with these faint sources.

Of the 410 variable sources in Table 2 only 114 have
astronomical identifications listed in the Simbad database. Of these sources, 77 (i.e. 67$\%$)
 can be
categorized as active galaxies (i.e. QSO's, AGN, BLL or Sy1). The next largest groups 
are those of X-ray sources (19), RR Lyrae stars (6)  and Novae (3). Although active
galaxies are the largest group
of $\it identified$ sources, it is highly probable that many of the remaining 296
uncategorized variable sources are of a stellar origin. This is mainly due to the poor coverage of 
astronomical identifications in the
SIMBAD database for faint stellar sources as opposed to that for active galaxies.

\subsection{RR Lyrae Stars}
GALEX has been shown to be a sensitive probe of the flux variability
observed towards RR Lyrae stars, since these stars can vary by up to
6 magnitudes at FUV wavelengths \citep{wheat05}. 
Using the known UV variable sources listed in the
GUVV-1 catalog, \cite{browne05} found that plots
of ($\it g$ - $\it r$) versus ($\it u$ - $\it g$)
SDSS  magnitudes revealed a segregation of stellar sources that
could be readily identified with RR Lyrae and $\delta$ Scuti-type stars. In Figure 5
we have produced a similar plot of ($\it g$ - $\it r$) versus ($\it u$ - $\it g$)
SDSS magnitudes for the present GUVV-2
set of variable sources listed in Table 2. The box drawn on
this Figure (i.e. 0.99 $<$ ($\it u$ - $\it g$) $<$ 1.28 and -0.20 $<$ ($\it g$ - $\it r$ ) $ < $ +0.31)
contains 36 sources, all of which are listed separately in Table 3. Only 2 of these sources are
previously known RR Lyrae stars (i.e. HL Her and V851 Her), and in
Figure 7(a) we show the magnitude-phase light curve for the newly discovered RR Lyrae star,
SDSS J081226.4+033320.1 that has a period of 0.555 days. This type
of UV observation, when used
in conjunction with ground-based ROTSE data,  can be
used to constrain Kurucz stellar atmosphere model parameters such as
stellar temperature and metallicity for RR Lyrae stars \citep{wheat05}.
Finally, we note that Table 2 also lists 4 previously known
RR Lyrae stars (with SIMBAD identifications) that are not shown in Figure 5. These 
stars, which have had their RR Lyrae status confirmed by other means, have not been included in Table 3 since their SDSS color magnitudes are not available.. 

\subsection{AGN and QSO's}
As previously stated, the largest group of known UV variable sources in Table 2 are active
galaxies. Quasars, blazars, BL Lac and Seyfert galaxies have long been known to exhibit
flux variability at all observed wavelengths from radio to gamma-rays,
lasting over time-scales of weeks, days or hours (micro-variability).
Flux variability studies, particularly at X-ray wavelengths, have been used to provide clues to the sizes and structure of the emission regions producing the observed variable level of radiation.
The non-thermal emission from the nucleii of active galaxies reveals itself as a flat UV-to-optical continuum spectrum, with the probable origin of the nuclear activity arising around
a supermassive black hole situated in the centre of the host galaxy which accretes gas from the host. During this process, gravitational binding energy is released, part of which is transformed into the UV radiation observed by GALEX. A typical UV spectrum of a low red-shift blazar is given in \cite{pian05}, with the majority of the flux being contained in the emission lines of SiIV, CIV and CIII 
superposed on an underlying continuum in
the GALEX FUV channel.
The GALEX NUV channel contains no strong emission lines with only UV continuum emission
being observed from these sources. 

Although we are presently unable to positively determine which of the sources listed in Table 2 are
true blazars, since the
number of presently known objects is small (especially so for the southern
galactic hemisphere), we believe it useful to produce a list of possible blazar candidates
for subsequent study and confirmation at other wavelengths.
Both \cite{seibert05} and \cite{bianchi07} have
attempted to isolate low red-shift QSO candidate sources in the GALEX
data archive from color-color
diagrams that use both UV and SDSS color magnitudes. In particular, plots of
(m$_{FUV}$ - m$_{NUV}$) versus (m$_{NUV}$ - $\it r$) colors can provide a very 
helpful discriminant between hot UV bright stars and low red-shift QSO's. In Figure 6 we
show such a color-color plot for all objects in Table 2 that possess  both an SDSS $\it r$ color
magnitude and which exhibit variability in both the FUV and NUV channels. Sources previously
identified as QSO's in SIMBAD are plotted as filled circles, and those without previous identification
are plotted as open circles. Following the work of  \cite{seibert05} and \cite{bianchi07}, it is
highly probable that sources lying within the region bounded by
$-0.5 < (FUV_{max} - NUV_{max}) < +1.5$
and (NUV$_{max}$ - $\it r$) $<$ 2.0 are low red-shift QSO's or
possibly blazars. We have identified these 35 objects
in Table 2 (see footnote to Table 2). This sample could possibly be contaminated by a few cataclysmic
variable stars but, as discussed in \cite{bianchi07}, such objects are quite rare compared to
the observed space density of QSO's.
Finally, we note the the
group of sources in Figure 6 that possess (FUV$_{max}$ - NUV$_{max}$) $>$ +2
are most likely to be variable main-sequence objects.

\subsection{Some Interesting GUVV-2 objects}
Of the 410 objects listed in Table 2, three sources stand out as being of particular
interest. These are (i) the contact binary system of SDSS J141818.97+525006.7,
(ii) the eclipsing dwarf nova system of IY UMa (SDSS J104356.72+580731.9)
and (iii) the highly variable unidentified source SDSS J104325.06+563258.1. 

In Figure 7(b) we show the NUV light curve for SDSS J141818.97+525006.7,  which
is a contact binary system listed in the catalog of \cite{gettel06}.
This type of highly
evolved system, often referred to as W Ursae Majoris (W UMa)
variables, consists of two very close binary stars whose outer atmospheric surfaces share
a common envelope. Noting that there are $\it two$ eclipses per cycle in the phased
UV light-curve, and we derive a period of 0.29064 d for this system which can be
compared with that of 0.29082 d obtained from visible data. Finally, we note that
the period for
this system is very close to that
of 0.27 d , which \cite{rucin07} cites as the maximum of the period distribution for all
known W UMa systems.

IY UMa is one
of the few known eclipsing dwarf nova systems of the SU UMa-type, consisting of a white dwarf 
which accretes matter via a Roche lobe flow from a late-type donor star \citep{uemura00}. 
Such systems typically undergo series of both short (days) and long (weeks)
outbursts. The GALEX data, shown
in Figure 8(a) represent the first UV
observations of this system, in which variability of $\sim$ 1 mag. has been detected over a set
of several observations spanning $\sim$ 1 year.
Previous visible spectrophotometric and
spectroscopic observations of this system 
have revealed both normal and super-outbursts which have been successfully modeled
by \cite{patterson00} and \cite{rolfe05}. 

Finally, in Figure 8(b) we show sets of NUV and FUV observations of the 
highly variable source SDSS J104325.06+563258.1 recorded with GALEX over a 24 month time-frame.
This source appears to possess FUV and NUV `quiescent'  magnitudes of $\sim$ 20 mag,
as measured during two observations some 450 days apart, followed by
two flare events which caused an increase
of $>$ 2 magnitudes. Although the temporal coverage of these outbursts is not
contiguous, the data do allow us to note that there was a rise of 3.7 mag in the NUV
in 40 days and on one occasion there was a drop of $\sim$ 2.0 mag in both the NUV and FUV
channels in only 21 hours.
Its SDSS {\it ugriz} magnitudes are all similar, suggesting
a flat spectral shape in the visible region, which would preclude its identification as a flaring
M-dwarf star. Although ground-based spectroscopic observations are required to provide
a firm identification for this source, it is probably a previously unknown distant dwarf nova system.

\subsection{Future Studies}
The great majority of the presently detected variable sources will require follow-up spectroscopic study
in order to reveal their true physical identity. However, it is clear from the 114 sources that already have
previous identifications that the GUVV-2 catalog will be a very useful tool in the study (and
indentification)
of the UV variability of active galaxies. 

All of our present variable source detections have been based on their
orbit-to-orbit variation in UV magnitude. The present study has not used the photon
data from each source to search for variability on time-scales of the order of seconds.
Previous studies of time-tagged photon data from 1800 GALEX images have
revealed short-term ($<$ 100 sec) UV flare events from 49 newly identified
dMe stars \citep{welsh07}. Future inspection of the photon data from the entire
GALEX data-base could enable studies of a variety of new source phenomenon that
includes the possible detection of stellar transits by Jupiter-sized exo-planets and the study of
short-term flare outbursts from cataclysmic variable systems. 

\begin{acknowledgments}
We gratefully acknowledge NASA's support for construction, operation,
and science analysis for the GALEX mission,
developed in cooperation with the Centre National d'Etudes Spatiales
of France and the Korean Ministry of
Science and Technology. We acknowledge the dedicated
team of engineers, technicians, and administrative staff from JPL/Caltech.
Financial support for this research was provided by NASA grant
NNG06GD17G through the GALEX Guest Investigator program. This publication makes use of data products from the SIMBAD database,
operated at CDS, Strasbourg, France, and the Sloan Digital Sky Survey.
\end{acknowledgments}

\begin{figure}
\center
{\includegraphics[height=8cm]{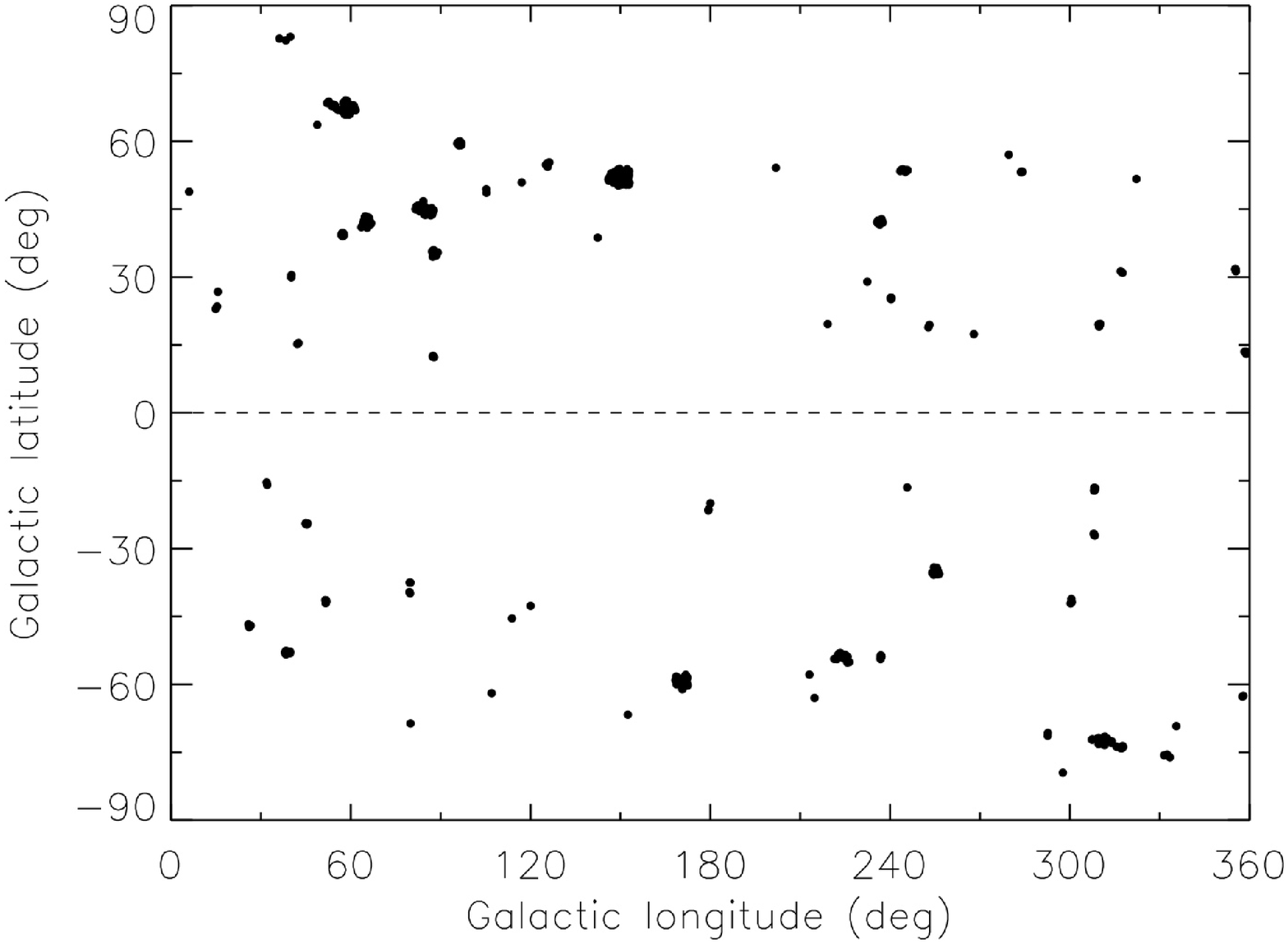}}
\caption{Distribution of the GUVV-2 catalogued sources on the sky.} 
\label{Figure 1}
\end{figure}

\begin{figure}
\center
{\includegraphics[height=6cm]{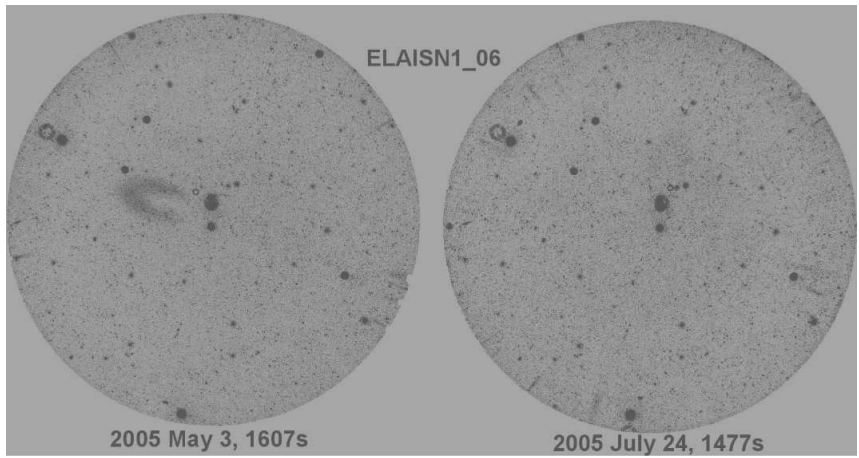}}
\caption{Two GALEX NUV images of the sky-field ELAIS-N1 recorded on 2005 May 3rd and 2005 July 24th. Note the fuzzy horseshoe-shaped object to the left of center of the May 3rd image that re-appears in the corresponding July 24th image with a greatly reduced intensity and different orientation angle to the rest of the image. This feature is caused by scattered light-leakage for bright objects lying just beyond the GALEX instrument field-of-view. } 
\label{Figure 2}
\end{figure}

\begin{figure}
\center
{\includegraphics[height=10cm]{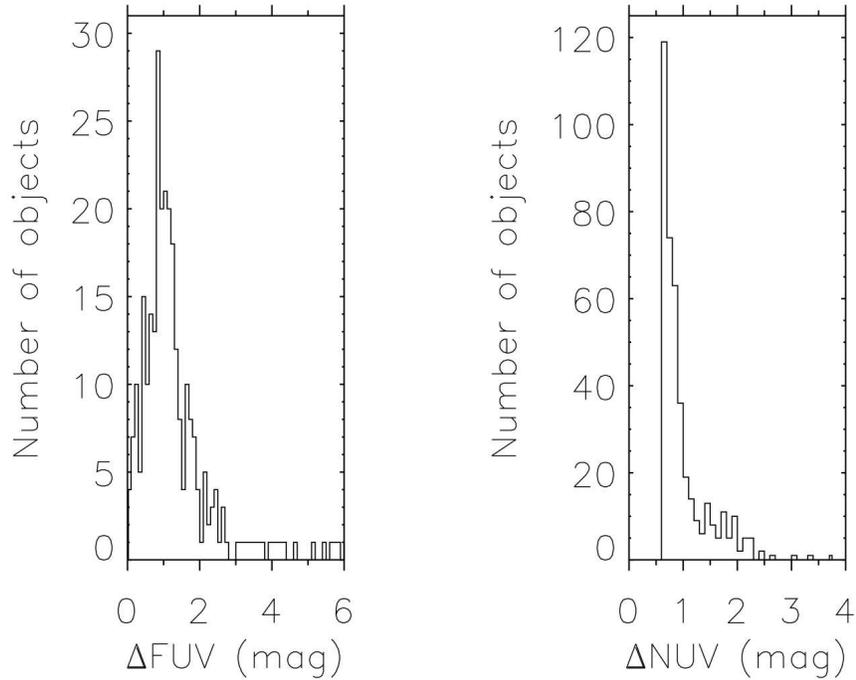}}
\caption{Plots of the maximum GALEX magnitudes, FUV$_{max}$ and NUV$_{max}$, as a function of the number of UV sources exhibiting such variability. } 
\label{Figure 3}
\end{figure}

\begin{figure}
\center
{\includegraphics[height=6cm]{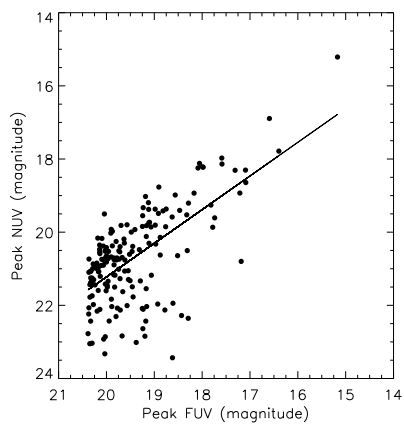}}
\caption{Plot of the GALEX  FUV peak magnitude vs. the peak NUV magnitude for the GUVV-2 sources listed in Table 2. The majority of sources lie within $\pm1$ mag of a straight line of slope +0.91.  } 
\label{Figure 4}
\end{figure}

\begin{figure}
\center
{\includegraphics[height=8cm]{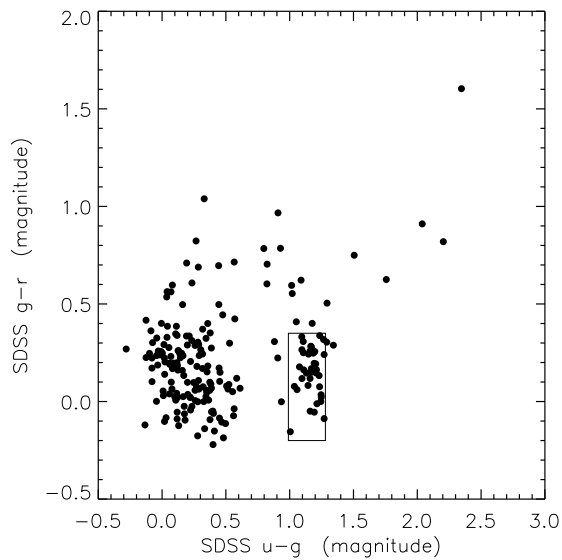}}
\caption{Plot of the SDSS ($\it g$ - $\it r$)  versus ($\it u$ - $\it g$) magnitudes for the GUVV-2 catalog sources. The 36 sources contained within the drawn box are most probably either RR Lyrae or $\delta$ Scuti-type variables.} 
\label{Figure 5}
\end{figure}

\begin{figure}
\center
{\includegraphics[height=8cm]{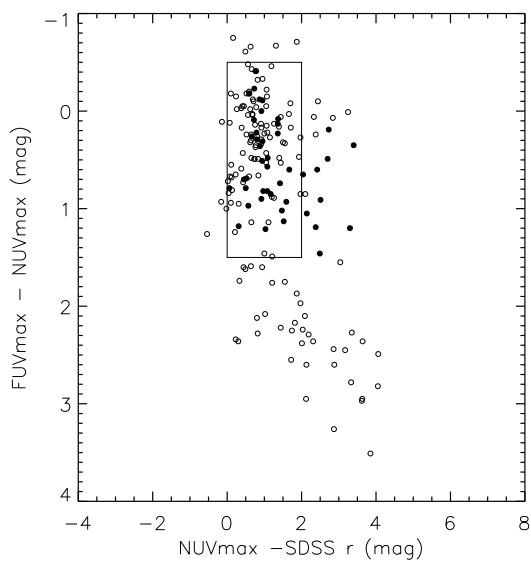}}
\caption{Plot of (FUV$_{max}$ - NUV$_{max}$) vs. (NUV$_{max}$ - SDSS $\it r$) for the variable sources in Table 2. Filled circles are sources previously identified as QSO's, open circles have no current identification. The open circle sources contained with the drawn box are most probably newly identified QSO's, and are marked as PQ (possible QSO's) in column 4 of Table 2.} 
\label{Figure 6}
\end{figure}

\begin{figure}
\center
{\includegraphics[height=7cm]{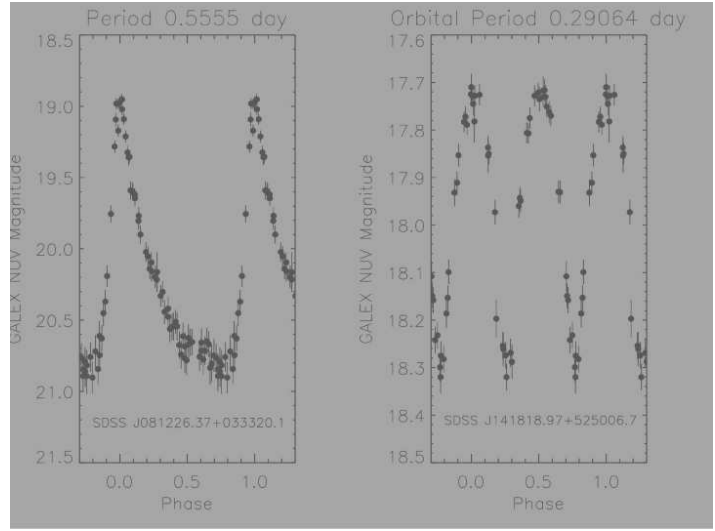}}
\caption{(a) Phased UV light-curve for the newly discovered RR Lyrae star, SDSS J081226.4+033320.1 which has an estimated period of 0.555 d . (b) The NUV light curve for the contact binary system SDSS J141818.97+525006.7,  which
has a derived UV period of 0.29064 d . } 
\label{Figure 7}
\end{figure}

\begin{figure}
\center
{\includegraphics[height=7cm]{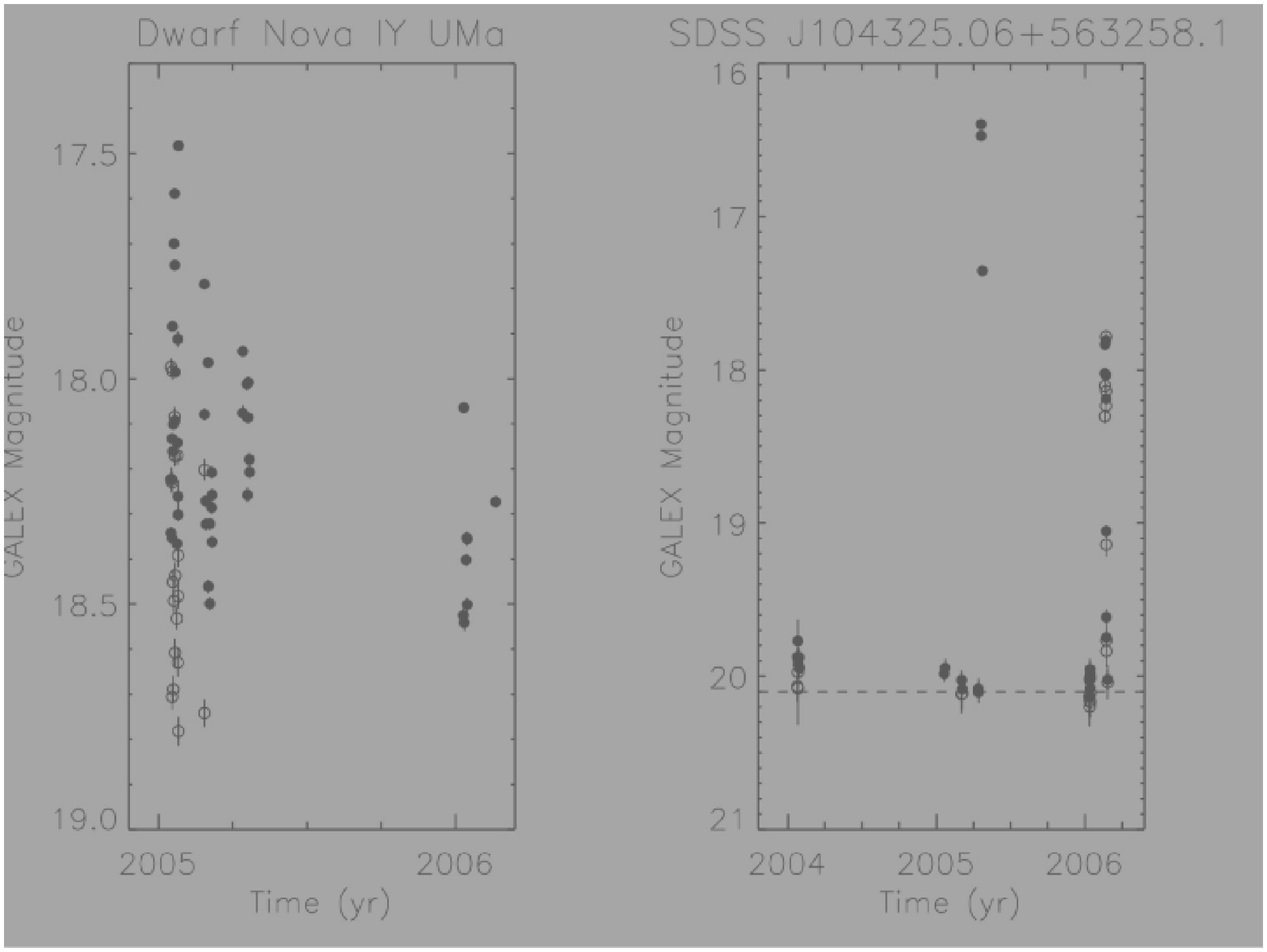}}
\caption{(a)Variation in FUV (open circles) and NUV (filled circles) magnitudes for the eclipsing dwarf nova system, IY UMa recorded at several epochs during a period of 12 months. (b) Unusually large variations in both FUV (open circles) and NUV (filled circles) magnitudes for the unidentified source, SDSS J104325.06+563258.1 recorded over a 2 year time-frame.} 
\label{Figure 8}
\end{figure}

 commands



\end{deluxetable}

\end{document}